\documentclass[english,aps,prl,superscriptaddress,reprint,longbibliography]{revtex4-1}
\usepackage[T1]{fontenc}
\usepackage{color}
\usepackage{amsmath}
\usepackage{amssymb}
\usepackage{graphicx}
\usepackage{wasysym}

\makeatletter
\usepackage{amsfonts}
\usepackage{graphicx}
\usepackage{xcolor}
\definecolor{darkblue}{rgb}{0,0,0.5} 
\usepackage{transparent}

\makeatother

\usepackage{babel}
\begin{document}

\title{Correlated photon dynamics in dissipative Rydberg media}

\author{Emil Zeuthen}
\email{zeuthen@nbi.ku.dk}
\affiliation{Niels Bohr Institute, University of Copenhagen, DK-2100 Copenhagen,
Denmark}
\affiliation{Institute for Theoretical Physics and Institute for Gravitational
Physics (Albert Einstein Institute), Leibniz Universit\"{a}t Hannover,
Callinstra{\ss}e 38, 30167 Hannover, Germany}

\author{Michael J. Gullans}
\affiliation{Joint Quantum Institute and Joint Center for Quantum Information
and Computer Science, National Institute of Standards and Technology
and University of Maryland, College Park, Maryland 20742, USA}
\author{Mohammad F. Maghrebi}
\affiliation{Joint Quantum Institute and Joint Center for Quantum Information
and Computer Science, National Institute of Standards and Technology
and University of Maryland, College Park, Maryland 20742, USA}
\affiliation{Department of Physics and Astronomy, Michigan State University, East Lansing, Michigan 48824, USA}
\author{Alexey V. Gorshkov}

\affiliation{Joint Quantum Institute and Joint Center for Quantum Information
and Computer Science, National Institute of Standards and Technology
and University of Maryland, College Park, Maryland 20742, USA}
\begin{abstract}

Rydberg blockade physics in optically dense atomic media under the conditions of electromagnetically induced transparency (EIT) leads 
to strong dissipative interactions between single photons.
We introduce a new approach to analyzing this challenging many-body
problem in the limit of large optical depth per blockade radius. In our approach, we
 separate the single-polariton EIT physics from Rydberg-Rydberg interactions in a serialized manner while using a hard-sphere
model for the latter, thus capturing the dualistic particle-wave nature of light as it manifests itself in dissipative Rydberg-EIT media.
Using this approach, we analyze the saturation behavior of the transmission
through one-dimensional Rydberg-EIT media in the regime of non-perturbative
dissipative interactions relevant to current experiments.  
Our model is able to capture the many-body dynamics of bright, coherent pulses through these strongly interacting media.  
We compare our model with available experimental data in this regime and find good agreement. 
We also analyze
a scheme for generating regular trains of single photons from continuous-wave input and derive its scaling behavior in the presence of imperfect single-photon EIT.
\end{abstract}

\maketitle

In optically dense atomic media, strong non-linear dissipative~\cite{Pritchard2010,Petrosyan11,Dudin2012,Peyronel2012,Maxwell2013,Gorshkov2013}  and dispersive~\cite{firstenberg13,bienias14,Maghrebi2015a} inter-photon interactions
at the single-quantum level can be engineered via interactions
of hybrid atom-photon excitations called Rydberg polaritons~\cite{Lukin2001,Comparat:10},  propagating due
to Electromagnetically Induced Transparency (EIT)~\cite{Fleischhauer2005}. Such photon-photon interactions can 
implement quantum gates between a pair of photons~\cite{Gorshkov2011,Shahmoon2011,Paredes-Barato2014};
experimental realizations in this area include single-photon phase
gates~\cite{Tiarks16} and transistors \cite{Tiarks14,Hofferberth14}.
These interactions between photons can also lead to the generation of anti-bunched \cite{Peyronel2012,Maxwell2014} and other non-classical \cite{Nielsen2010,Pohl2010,Bariani2012,Pritchard2012,Stanojevic2012,Gorshkov2013,Grankin2014,Maghrebi2015b} states of light.

The rich physics of dissipative Rydberg-EIT media and their potential as building blocks in quantum information science make urgent
the challenge to understand the underlying quantum many-body dynamics. Whereas
the few-photon theory has been established~\cite{Peyronel2012},
a tractable approach for analyzing the high-intensity sector of interest
to ongoing experiments \cite{Peyronel2012,Hofferberth16} has thus far not been available.
In this Letter, 
we overcome these shortcomings by constructing a model  for the many-body dynamics of
one-dimensional dissipative Rydberg-EIT media.  
We compare our model with  experimental data from Ref.~\cite{Peyronel2012} and find good agreement. 
We then show how this system can generate a regular train of single photons from uniform coherent-state input.  Such photon trains have wide utility in applications such as boson sampling \cite{Motes2014}, quantum key distribution \cite{zoller05}, and sub-shot-noise imaging of low-absorption samples \cite{brida10}.  At a more fundamental level, the emergence of a strongly correlated state of photons from a classical input is an interesting case study for emergent behavior in non-equilibrium quantum many-body systems.

\begin{figure}
\begin{centering}
\includegraphics[width=1\columnwidth]{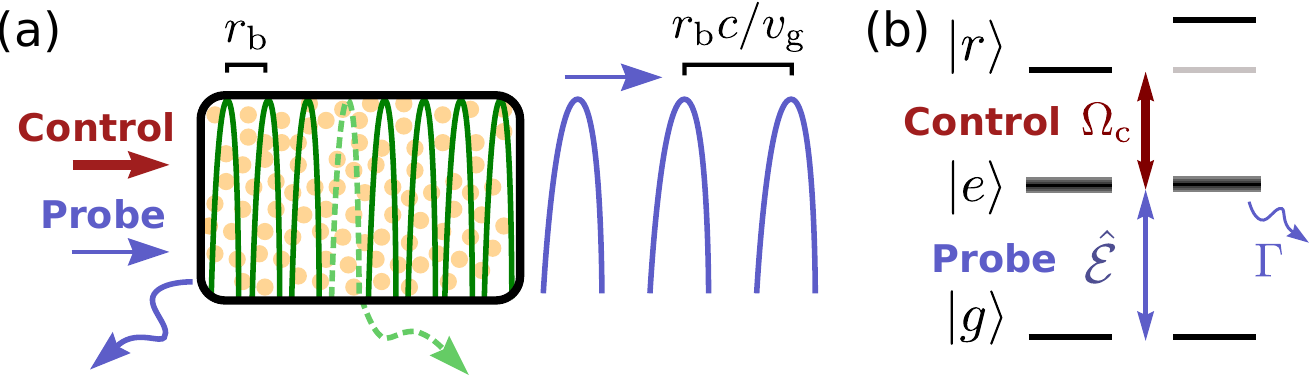}
\par\end{centering}
\caption{Dissipative Rydberg blockade in a one-dimensional atomic EIT medium.
(a) Dynamics resulting from high-intensity CW probing according to
naive expectation. The one-dimensional wave train of polaritons and
outgoing photons is indicated. (b) Level diagram for the atoms leading
to a dissipative blockade. Ground and Rydberg levels, $|g\rangle$
and $|r\rangle$, are long-lived compared to the lossy excited level
$|e\rangle$ with decay rate $2\Gamma$. Outside the blockade
radius of Rydberg polaritons, an incoming photon enjoys EIT transmission
(left set of energy levels). For atoms within the blockade region
of a polariton, the Rydberg level $|r\rangle$ is shifted out of resonance
with respect to the classical drive $\Omega_{{\rm c}}$ causing an
incoming photon to scatter out of the excited state $|e\rangle$.\label{fig:System-and-levels}}
\end{figure}

In a simplistic picture of the high-intensity dynamics of such a medium
[Fig.~\ref{fig:System-and-levels}(a)], a bright continuous-wave (CW) probe field produces
a train of Rydberg polaritons (green peaks). 
Via Rydberg-Rydberg (R-R) interactions, one polariton destroys lossless EIT-propagation conditions for other nearby polaritons. As a result, inside the medium, the polaritons are separated by their blockade
radius $r_{\text{b}}$ due to the scattering of ``superfluous''
photons at the entrance (purple wavy arrow) out of the forward-propagating
mode. In turn, the polaritons exit the medium
as a train of single photons spaced by the decompressed blockade radius
$r_{\text{b}}c/v_{\text{g}}$ ($v_{\text{g}}$ is the polariton
group velocity and $c$ is the speed of light).
However, polaritons may decay (green dashed arrow) due to the finite
width of the EIT window~\cite{Peyronel2012},
whereby the observed anti-bunching feature will have an extent
which significantly exceeds the blockade time $\tau_{\text{b}}\equiv r_{\text{b}}/v_{\text{g}}$.
This is due to the spectral features of width $\sim1/\tau_{\text{b}}$
generated by the blockade exceeding the bandwidth of the EIT transmission
window. Consequently, these features are washed out due to EIT filtering
as the polaritons propagate through the medium.
We include in our model this detrimental \emph{single-polariton} EIT effect, that sets the limits of control for Rydberg-EIT media, but has been ignored in existing theories for the high-intensity regime.

Rydberg-EIT blockade experiments have been performed in both free
space~\cite{Pritchard2010,Dudin2012,Peyronel2012,Maxwell2013,Hofmann2013}
and intra-cavity~\cite{Boddeda2016} settings. Focusing here on
the former, a number of theoretical approaches to the dissipative
blockade can be found in the literature: Theories based on semi-classical
weak probing~\cite{Sevincli2011} and quantum super atoms~\cite{Petrosyan2011,Liu2014,Garttner2014},
respectively, have both successfully predicted the experimental transmission
data of Ref.~\cite{Pritchard2010}. Also accounting for the quantum
nature of light, Ref.~\cite{Gorshkov2011} analyzed the case of two
strongly interacting photons. Here it was demonstrated that dissipative
Rydberg-EIT media subject to copropagating input photons will produce
an output field exhibiting the avoided volume associated with blockade,
an effect that may serve as the basis of a single-photon filter in the limit of large optical depth per blockade radius $d_{\text{b}}\gg1$~\cite{Gorshkov2013}. 

Our model has two main advantages over previous work in this direction~\cite{Gorshkov2013}:
Firstly, by allowing for input pulses exceeding the blockade time
$\tau_{\text{b}}$, we can analyze the CW limit of operation.
Secondly, we incorporate the fact that single-polariton EIT decay
releases the blockade, thereby allowing for the formation of a new
polariton. 

To set up the model, we start by considering the
limit of perfect single-polariton EIT, $d_{\text{b}}\gg1$,
allowing us to discuss the R-R interactions independently. These arise
in an ensemble of 3-level atoms in the ladder configuration [Fig.~\ref{fig:System-and-levels}(b)],
in which the van der Waals potential associated with a Rydberg excitation
$|r\rangle$ tunes neighboring atoms out of the EIT condition, thus
activating the dissipation channel of the excited state
$|e\rangle$. For simplicity, 
we replace the potential by a hard-sphere potential of radius $r_{\text{b}}$ such that 
inside this region the photons immediately
scatter, whereas outside they are unaffected.

\begin{figure}
\centering
\def\svgwidth{\columnwidth}
\includegraphics[width=1\columnwidth]{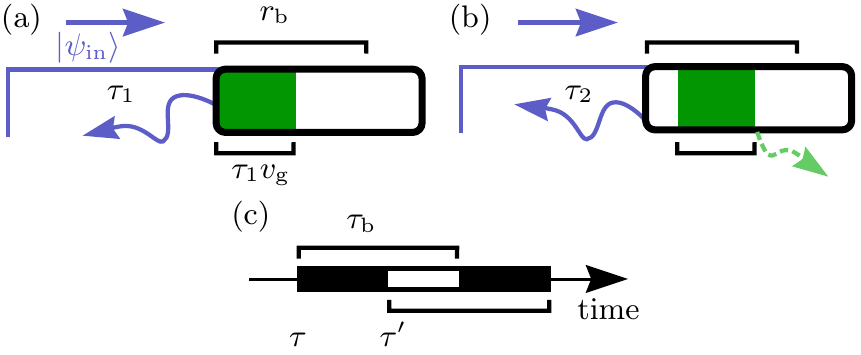}

\caption{Formation of the polariton wave function by projection 
of an incoming pulse in state $|\psi_{\text{in}}\rangle$ according to the
hard-sphere model in the limit of perfect single-polariton EIT. (a)
An incoming probe photon is scattered (purple wavy arrow) near the
beginning of the medium at time $\tau_{1}$ 
thereby creating a polariton pulse (green rectangle)
in the medium. (b) This pulse propagates
further into the medium until at a time $\tau_{2}$ a second
probe photon scatters, but since (in this instance) the polariton
could not have left the first $r_{\text{b}}$ of the medium, no additional
projection of the polariton wave function ensues (see \cite{SM} for
an extended discussion). The short duration of the polariton leads to
EIT losses (green dashed arrow). (c) Geometrical
derivation of polariton coherence: The overlap of the blockade regions during arrival times $\tau$ and $\tau'$ is indicated by shading. The ``forbidden'' region, during which scattering events
would ruin the superposition between $\tau$ and $\tau'$, is shown in black (here assuming $|\tau - \tau'|\leq \tau_{\text{b}}$).
\label{fig:Successive-projections}}
\end{figure}

Within this hard-sphere model, the action of the Rydberg
medium on a pulsed input can be understood as sketched in Fig.~\ref{fig:Successive-projections}:
If a photon successfully enters
the medium, it forms a polariton propagating without loss at speed
$v_{\text{g}}$. If another photon enters at time
$\tau_{1}<\tau_{\text{b}}$ after the polariton was created, it will scatter
out of the forward-propagating mode. 
As a result, 
the environment effectively projects the wave function of the polariton
to be localized within the arrival time interval $[\tau_{1}-\tau_{\text{b}},\tau_{1}]$
[see Fig.~\ref{fig:Successive-projections}(a)]. As long as the polariton
is within the first blockade radius of the entrance, subsequent scattering
events will not further localize the polariton [see Fig.~\ref{fig:Successive-projections}(b)];
hence, only the single-polariton physics plays a role during the propagation
through this region. 
Once the polariton has propagated a distance
$r_{\text{b}}$ into the medium, it can no longer cause scattering
of incoming photons and a new polariton may be formed at the entrance
of the medium~\footnote{Generally, this will lead to entanglement between polaritons and,
hence, between the outgoing photons as discussed in the Supplemental
Material.}. When a polariton reaches the rear of the medium, it will map back
onto the outgoing optical field after undergoing spatial decompression by a factor of $c/v_{\text{g}}$. 

\textit{Dissipative theory at high intensity.---}We will now apply these ideas to 
analyze two scenarios involving
dissipative Rydberg-EIT media in the one-dimensional limit, where the
transverse spot size of the impinging light fields is small compared
to $r_{\text{b}}$. As our first application, we  consider
the transmission behavior for a medium subject to CW probe and control
fields as a function of the input rate $\mathcal{R}_{\text{in}}$
of the probe. 

Preliminarily, we derive the output rate
in absence of single-polariton EIT decay, i.e., only considering the
R-R interaction. The input rate $\mathcal{R}_{\text{in}}$ splits
into two fractions, the rate of photons that make it through the
medium $\mathcal{R}_{\text{out}}$ and a rate of R-R scattered photons,
$\mathcal{R}_{\text{out}}\tau_{\text{b}}\mathcal{R}_{\text{in}}$~\footnote{We neglect finite-pulse corrections that arise from a finite averaging time.}:
\begin{equation}
\mathcal{R}_{\text{in}}=\mathcal{R}_{\text{out}}+\mathcal{R}_{\text{out}}\tau_{\text{b}}\mathcal{R}_{\text{in}};\label{eq:R-in_fractions-perf-EIT}
\end{equation}
$\tau_{\text{b}}\mathcal{R}_{\text{in}}$ is the average number
of photons scattered by each Rydberg polariton, while $\mathcal{R}_{\text{out}}$
is the rate of periods where the medium is blockaded. From Eq.~(\ref{eq:R-in_fractions-perf-EIT}),
we find the desired result~\cite{Muller1974}
\begin{equation}
\mathcal{R}_{\text{out}}=\frac{1}{\tau_{\text{b}}+1/\mathcal{R}_{\text{in}}}.\label{eq:R-out_perf-EIT}
\end{equation}
As expected, the output rate $\mathcal{R}_{\text{out}}$ is upper-bounded by both the input rate $\mathcal{R}_{\text{in}}$ and the inverse blockade time
$1/\tau_{\text{b}}$, with each bound achievable in the limit of weak blockade $\tau_{\text{b}} \rightarrow 0$ and high input rate $\mathcal{R}_{\text{in}} \rightarrow \infty$, respectively.  

We now turn to the more realistic case of imperfect single-polariton
EIT, in which we must account for the finite transmission of polaritons
that do not fit within the EIT window. If a polariton is EIT-scattered
within the first blockade radius $r_{\text{b}}$ of the medium, its
blockading effect ceases, hence allowing for the creation of a new polariton
at the input of the medium. Therefore the average blockade time per
polariton $\bar{\tau}_{\text{b}}$ is upper-bounded by that of a polariton
that makes it through the first $r_{\text{b}}$ of the medium, $\bar{\tau}_{\text{b}}\leq\tau_{\text{b}}$.
Introducing the average single-polariton EIT transmission $\bar{\eta}_{\text{EIT}}(L)$
through a medium of length $L\geq r_{\text{b}}$ and the average Rydberg
formation rate $\mathcal{R}_{\text{Rf}}$, we  decompose the
input rate in the spirit of Eq.~(\ref{eq:R-in_fractions-perf-EIT}),
\begin{equation}
\mathcal{R}_{\text{in}}=\bar{\eta}_{\text{EIT}}(L)\mathcal{R}_{\text{Rf}}+[1-\bar{\eta}_{\text{EIT}}(L)]\mathcal{R}_{\text{Rf}}+\mathcal{R}_{\text{Rf}}\bar{\tau}_{\text{b}}\mathcal{R}_{\text{in}},\label{eq:R-in_fraction-imperf-EIT}
\end{equation}
where on the right-hand side the first term is the output rate, $\mathcal{R}_{\text{out}}\equiv\bar{\eta}_{\text{EIT}}(L)\mathcal{R}_{\text{Rf}},$
the second term is the rate of polariton EIT decay, and the third
term is the rate of R-R scattering events. Introducing the output
rate $\mathcal{R}_{\text{out}}$ into Eq.~(\ref{eq:R-in_fraction-imperf-EIT})
and solving for this quantity, we find
\begin{equation}
\mathcal{R}_{\text{out}}=\frac{\bar{\eta}_{\text{EIT}}(L)}{\bar{\tau}_{\text{b}}+1/\mathcal{R}_{\text{in}}},\label{eq:R-out_imperf-EIT}
\end{equation}
generalizing Eq.~(\ref{eq:R-out_perf-EIT}). Equation~(\ref{eq:R-out_imperf-EIT})
shows that EIT decay alters the transmission not only by the trivial
damping prefactor $\bar{\eta}_{\text{EIT}}$, but also by decreasing
the effective blockade time $\bar{\tau}_{\text{b}}\leq\tau_{\text{b}}$.
Moreover, since both $\bar{\eta}_{\text{EIT}}$ and $\bar{\tau}_{\text{b}}$
are likely to decrease with increasing $\mathcal{R}_{\text{in}}$,
the transmission curve (\ref{eq:R-out_imperf-EIT}) need not be a monotonic function of $\mathcal{R}_{\text{in}}$.

To evaluate Eq.~(\ref{eq:R-out_imperf-EIT}), we estimate 
$\bar{\eta}_{\text{EIT}}(L)$ and $\bar{\tau}_{\text{b}}$ using the
idea of projection-free propagation discussed in connection with Fig.~\ref{fig:Successive-projections}(a,b).
This allows a serialized treatment: We take the temporal extent $\tau$
of a polariton to be defined by the first R-R event after its formation
and average over the Poisson distribution for the arrival times of
photons. The associated EIT transmission is then modeled by that of
a square pulse of duration $\tau$ subjected to Gaussian filtering,
yielding the following expressions~\cite{SM},
\begin{gather}
\bar{\eta}_{\text{EIT}}(l)=\exp\left([\tau_{\text{EIT}}(l)\mathcal{R}_{\text{in}}]^{2}\right)\text{erfc}\left(\tau_{\text{EIT}}(l)\mathcal{R}_{\text{in}}\right),\label{eq:eta_EIT-bar}\\
\frac{\bar{\tau}_{\text{b}}}{\tau_{\text{b}}}=\int_{0}^{r_{\text{b}}}\frac{dl}{r_{\text{b}}}\bar{\eta}_{\text{EIT}}(l)=\frac{\bar{\eta}_{\text{EIT}}(r_{\text{b}})+\sqrt{\frac{4}{\pi}}\tau_{\text{EIT}}(r_{\text{b}})\mathcal{R}_{\text{in}}-1}{[\tau_{\text{EIT}}(r_{\text{b}})\mathcal{R}_{\text{in}}]^{2}},\label{eq:eta-tau-bar_result}
\end{gather}
where the full blockade time is $\tau_{\text{b}}=d_{\text{b}}/(2\gamma_{\text{EIT}})$,
the filtering time $\tau_{\text{EIT}}(l)\equiv\sqrt{d_{\text{b}}l/r_{\text{b}}}/\gamma_{\text{EIT}}$
and $\gamma_{\text{EIT}}\equiv\Omega_{\text{c}}^{2}/\Gamma$ is the
single-atom EIT linewidth in terms of the control field Rabi frequency
$\Omega_{\text{c}}$ and the halfwidth $\Gamma$ of the intermediate
level~\footnote{$\tau_{\text{EIT}}$ is the inverse Gaussian width of the EIT amplitude filtering $\propto \text{exp}(-\omega^2 \tau_{\text{EIT}}^{2}/2)$ and the convention for $\Omega_{\text{c}}$ is set by $\partial_{t} \hat{P} = i \Omega_{\text{c}} \hat{S}$, where $\hat{P}$ and $\hat{S}$ are the polarizations between \unexpanded{$|g\rangle\text{--}|e\rangle$} and \unexpanded{$|g\rangle\text{--}|r\rangle$} transitions, respectively.}; taken together with these definitions, Eqs.~(\ref{eq:eta_EIT-bar},\ref{eq:eta-tau-bar_result})
determine the output rate (\ref{eq:R-out_imperf-EIT}).

\begin{figure}
\begin{centering}
\begin{picture}(320,165) 
\put(0,0){\includegraphics[width=1\columnwidth]{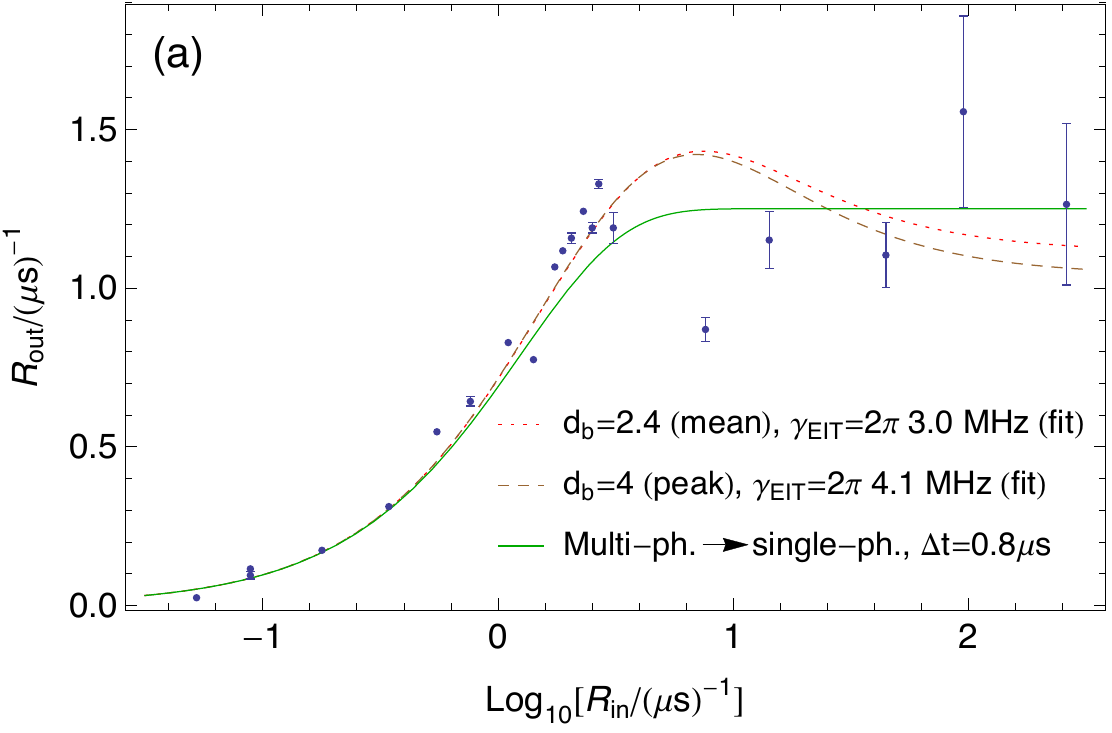}}
\color{white}{\put(40,150){\circle*{14}}}
\end{picture}
\end{centering}
\caption{Transmission through a dissipative Rydberg-EIT medium. Output
rate $\mathcal{R}_{\text{out}}$ as function of input rate $\mathcal{R}_{\text{in}}$.
The data from Ref.~\cite{Peyronel2012} is compared to plots of Eq.~(\ref{eq:R-out_imperf-EIT})
for two fixed values of $d_{\text{b}}$ corresponding to mean (red
dotted) and peak (brown dashed) values of the Gaussian atomic density distribution
in the experiment with axial spread $\sigma_{\text{ax}}$ and
effective length $L=4.2\times\sigma_{\text{ax}}$. The value for the
single-atom EIT linewidth in the experiment was $\gamma_{\text{EIT}}=2\pi\times7.5\text{ MHz}$.
Also plotted (solid green curve) is the rate resulting from converting
all multi-photon events within successive time intervals $\Delta t=0.8\,\mu\text{s}$
into single-photon events $\mathcal{R}_{\text{out}}=(\Delta t)^{-1}(1-e^{-\mathcal{R}_{\text{in}}\Delta t})$.
The latter was presented
alongside the data in Ref.~\cite{Peyronel2012}.\label{fig:R-out-comparison}}
\end{figure}

We plot Eq.~(\ref{eq:R-out_imperf-EIT}) in Fig.~\ref{fig:R-out-comparison}
using the experimental parameters of Ref.~\cite{Peyronel2012} and
compare to the data and the alternative theoretical curve presented
therein (however, $\gamma_{\text{EIT}}$ was obtained by fitting Eq.~(\ref{eq:R-out_imperf-EIT})
to the data resulting in values within a factor of 2 from that cited
in Ref.~\cite{Peyronel2012}). Within the error bars of the data, Eq.~(\ref{eq:R-out_imperf-EIT})
is seen to match the data equally well as the theoretical curve proposed
in Ref.~\cite{Peyronel2012}. Crucially, however, the physics implicit
in the latter functional form is at odds with that of the Rydberg
blockade: It puts no lower limit on the spacing between output photons,
and the time-scale $\Delta t$ that must be assumed to fit the data
is an order of magnitude larger than $\tau_{\text{b}}$ calculated
from experimental parameters. Eq.~(\ref{eq:R-out_imperf-EIT}) predicts
the following two features of the transmission curve: Firstly, a finite
asymptote for the limit of very high intensity, $\mathcal{R}_{\text{in}}\rightarrow\infty$;
while in this limit the average transmission probability goes
to zero, $\bar{\eta}_{\text{EIT}}(L)\rightarrow0$, the effective
blockade time will likewise decrease, $\bar{\tau}_{\text{b}}\rightarrow0$,
thereby increasing the polariton formation rate in a manner such that $\mathcal{R}_{\text{out}}$ in Eq.~(\ref{eq:R-out_imperf-EIT})
remains finite. Secondly, Eq.~(\ref{eq:R-out_imperf-EIT}) in general
exhibits a local maximum as in Fig.~\ref{fig:R-out-comparison}.

\textit{Generation of a single-photon train.---}We will now show that, in the limit  $d_{\text{b}}\gg 1$, a CW probe gets converted into a regular pulse train of single photons [see
 Fig.~\ref{fig:System-and-levels}(a)]. The blockade sets a lower limit $\tau_{\text{b}}$ to the temporal
separation of output photons leading to anti-bunching. To achieve
regularity, we must also impose an upper bound on the (average) separation.
This can be ensured by a sufficiently large input rate, but only insofar
as single-polariton EIT decay remains rare. Such a
decay event will terminate the regularity of the pulse train (creating
a ``domain wall''), effectively resetting the process. These considerations
imply that an optimum input rate exists as a trade-off between the
input photons not being too far apart (on average) while keeping single-polariton
EIT decay at a perturbative level. In this regime, we may 
take $\bar{\tau}_{\text{b}}\approx\tau_{\text{b}}$ and estimate $\bar{\eta}_{\text{EIT}}$
by filtering the correlation functions $G^{(1)}(\tau;\tau')=\langle\hat{\mathcal{E}}^{\dagger}(\tau)\hat{\mathcal{E}}(\tau')\rangle$ produced
by the idealized R-R interaction [see Fig.~\ref{fig:Successive-projections}(a,b)] 
(as was checked by numerical simulations~\cite{SM}). 

\begin{figure}
\begin{centering}
\includegraphics[width=1\columnwidth]{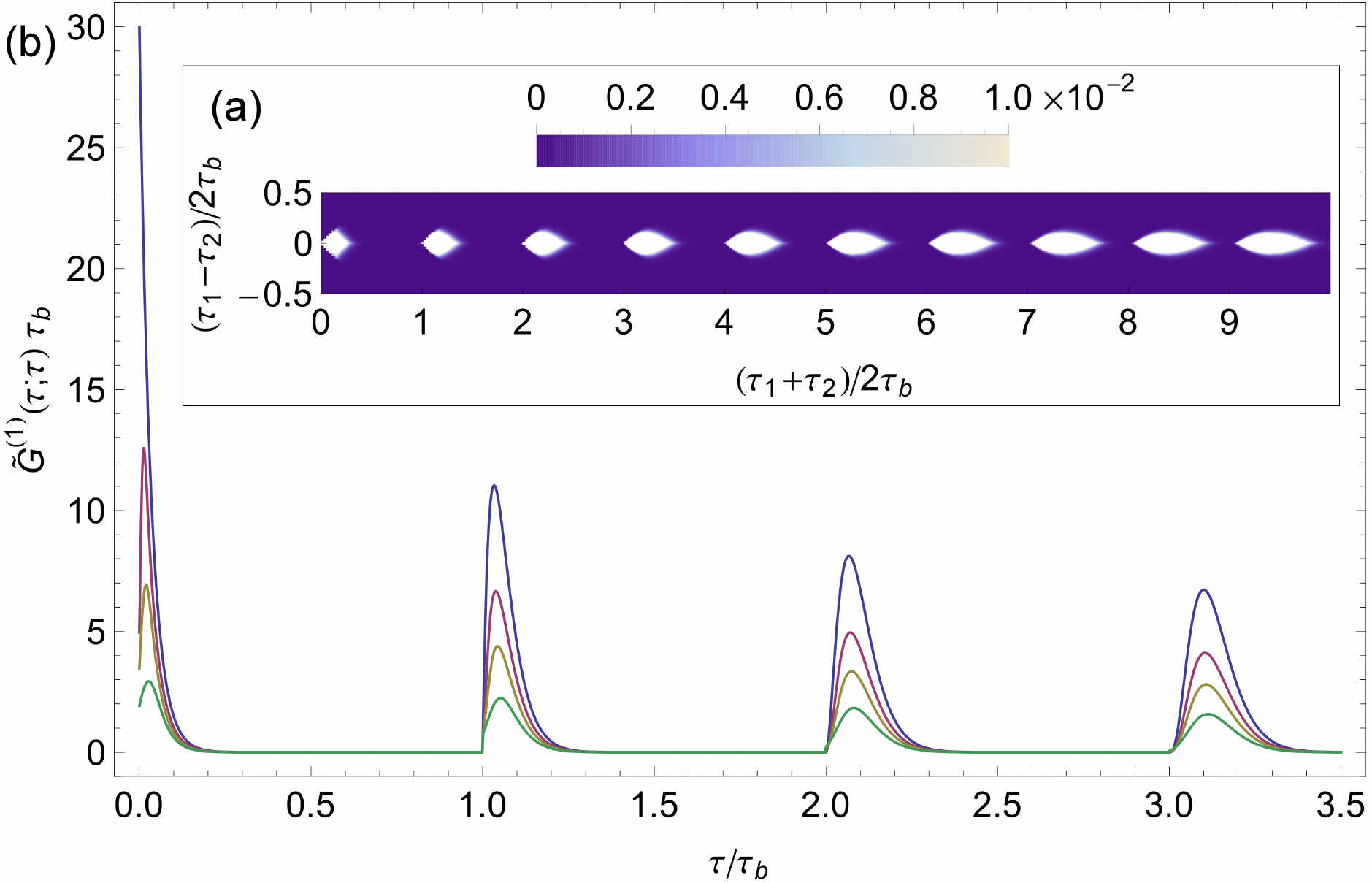}
\par\end{centering}
\noindent \centering{}\caption{(a) $G^{(1)}(\tau_{1};\tau_{2})$ {[}in units of $1/\tau_{\text{b}}${]}
for coherent square-pulse input entering the medium at $\tau=0$ in
the limit of perfect single-polariton EIT, Eq.~(\ref{eq:G1_off-diag_coherent-ignore-edge})
with $\mathcal{R}_{\text{in}}=30/\tau_{\text{b}}$ (values $>10^{-2}$
are plotted as white for illustrative purposes). (b) Diagonal elements
of the EIT-filtered correlation function $\tilde{G}^{(1)}(\tau;\tau)$
for EIT width $1/\tau_{\text{EIT}}$, where $\tau_{\text{EIT}}\mathcal{R}_{\text{in}}\in\{0,1/4,1/2,1\}$
(in order of decreasing amplitude: purple, magenta, yellow, green).\label{fig:G1_offdiag-coherent}}
\end{figure}

We first derive the diagonal element $G^{(1)}(\tau;\tau)$ by observing
that its value at a time $\tau$ after the onset of the pulse can
have contributions from at most $\left\lceil \tau/\tau_{\text{b}}\right\rceil $
polaritons per the hard-sphere ansatz. The contribution from each
polariton follows from straightforward integration of the Poisson
distribution; we then find ($\tau\geq0$)~\cite{SM}
\begin{equation}
G^{(1)}(\tau;\tau)=\sum_{j=0}^{\left\lfloor \tau/\tau_{\text{b}}\right\rfloor }\mathcal{R}_{\text{in}}e^{-\mathcal{R}_{\text{in}}(\tau-j\tau_{\text{b}})}\frac{[\mathcal{R}_{\text{in}}(\tau-j\tau_{\text{b}})]^{j}}{j!} \label{eq:G1-diag-Poisson}
\end{equation}
(the argument can be extended to obtain higher-order correlation functions~\cite{SM}).
This function is plotted in Fig.~\ref{fig:G1_offdiag-coherent}(b)
(top blue curve) in the high-intensity regime in which the statistical
uncertainty in the creation time of polaritons is small (relative
to $\tau_{\text{b}}$); i.e., when only the term $j=\left\lfloor \tau/\tau_{\text{b}}\right\rfloor$
in Eq.~(\ref{eq:G1-diag-Poisson}) contributes significantly.

The first-order correlation function $G^{(1)}(\tau;\tau')$
quantifies the quantum coherence between having a Rydberg excitation
at different times. This can be related to the diagonal elements
$G^{(1)}(\tau'';\tau'')$ by noting that $G^{(1)}(\tau;\tau')$
is the probability density for the creation of a polariton at time
$\tau_{<}\equiv\min\{\tau,\tau'\}$ multiplied by the probability
that no scattering events occur that allow the environment to distinguish
whether the polariton was created at $\tau$ or $\tau'$.
From Fig.~\ref{fig:Successive-projections}(c)
we see that this requires that no photons impinge on the medium for
intervals of combined duration $2|\tau-\tau'|$~\footnote{Assuming $\tau,\tau'\leq\tau_{\text{p}}-\tau_{\text{b}}$, where $\tau_{\text{p}}$
is the pulse duration.}, which is all the information needed for Poisson-distributed input (here focusing on $|\tau-\tau'|\leq\tau_{\mathrm{b}}$ relevant for the limit of interest, $d_{\textrm{b}} \gg 1$, in which $\tau_{\text{EIT}} \ll \tau_{\text{b}}$).
Hence ($|\tau-\tau'|\leq\tau_{\mathrm{b}}$)
\begin{equation}
G^{(1)}(\tau;\tau')=G^{(1)}(\tau_{<};\tau_{<})e^{-2\mathcal{R}_{\text{in}}|\tau-\tau'|},\label{eq:G1_off-diag_coherent-ignore-edge}
\end{equation}
where the diagonal elements of $G^{(1)}$ are given by Eq.~(\ref{eq:G1-diag-Poisson}).
Equation (\ref{eq:G1_off-diag_coherent-ignore-edge}) is plotted in Fig.~\ref{fig:G1_offdiag-coherent}(a).
By convolving Eq.~(\ref{eq:G1_off-diag_coherent-ignore-edge}) with
a Gaussian filter function corresponding to a medium of length $L=r_{\text{b}}$, we estimate the effect of EIT decay, obtaining
the intensity profile $\tilde{G}^{(1)}(\tau;\tau)$ plotted in Fig.~\ref{fig:G1_offdiag-coherent}(b).

Equation~(\ref{eq:G1-diag-Poisson}) captures how the accumulated randomness in the photon
arrival times determines the spread in the
formation times of polaritons.
To obtain a parameter condition ensuring regularity of the pulse train
for the first $N_{\text{loc}}$ peaks of $G^{(1)}(\tau;\tau')$, we demand
that the width $(\delta t)_{N}$ of the widest peak in the train obeys
$(\delta t)_{N}\leq\beta\tau_{\text{b}}$ for some fraction $\beta$
of $\tau_{\mathrm{b}}$, yielding the requirement ($N_{\text{loc}}\gg1$)
$\mathcal{R}_{\text{in}}\apprge\sqrt{N_{\text{loc}}}/(\beta\tau_{\text{b}})$~\cite{SM}. 
On the other hand, if we can tolerate an EIT loss fraction of at most
$\epsilon=1-\bar{\eta}_{\text{EIT}}$, filtering Eq.~(\ref{eq:G1_off-diag_coherent-ignore-edge})
yields an upper bound (in the limit
$\tau_{\text{EIT}}\ll1/\mathcal{R}_{\text{in}},\tau_{\text{b}}$), 
$\mathcal{R}_{\text{in}}\apprle 1/(N_{\text{EIT}}\tau_{\text{EIT}})$, 
where $N_{\text{EIT}}\sim1/\epsilon$ is the average pulse train length
allowed by the EIT filter. Balancing these two considerations by setting
$N=N_{\text{loc}}=N_{\text{EIT}}$ and seeking its maximal value $N_{\text{max}}$
that permits the simultaneous fulfillment of the above requirements, we find
\begin{equation}
N_{\text{max}}\sim\sqrt[3]{\frac{\pi\beta^{2}}{128\ln2}d_{\text{b}}},\label{eq:N-max_result}
\end{equation}
having used $\tau_{\text{b}} = d_{\text{b}}/(2\gamma_{\text{EIT}})$
and $\tau_{\text{EIT}}=\sqrt{d_{\text{b}}}/\gamma_{\text{EIT}}$.
For this expression to reach $N_{\text{max}}\apprge1$ requires $d_{\text{b}}\apprge100$
for $\beta\approx1/2$ and hence we find the generation of single-photon pulse trains
to demand very large $d_{\text{b}}$.
The scaling behavior \eqref{eq:N-max_result} resulting from the limiting effects identified here may be improved by directing the scattered light into well defined channels using cooperative light emission \cite{Murray2017}.

In conclusion, we have proposed a new model for analyzing the many-body
physics of the dissipative Rydberg blockade in extended one-dimensional
EIT media. The work presented here may serve as a guide
for the rigorous derivation of effective many-body theories~\cite{Gullans2016}.
In this regard, we remark that ongoing work on numerically simulating Rydberg-EIT media using matrix product states yields results which are in qualitative agreement with the present work~\cite{Manzoni2017,DouglasInPrep}.

One can envision several extensions of our approach: Storage
and retrieval operations in Rydberg media can be modeled by translating
time-varying control fields into a time-dependent blockade time $\tau_{\text{b}}$
of hard-sphere Rydberg polaritons. In three-dimensional Rydberg media, multi-body
Rydberg scattering events can take place when the blockade regions
of transversely spaced polaritons overlap. It would also be intriguing to consider whether our protocol for the generation of photon trains can be modified, perhaps by employing cooperative resonances in regular atomic arrays \cite{Shahmoon2017}, to turn the photon trains into time crystals \cite{wilczek12b,shapere12} that keep their periodicity for times that are exponentially long in the input intensity. 

\begin{acknowledgments}
E.\,Z.~is grateful to J.\,Taylor and the Joint Quantum Institute for hosting him during the time when
this project was conceived. We thank the Vuleti\'{c}-Lukin group
for providing their experimental data. 
We thank Q.-Y.\,Liang, J.\,Thompson, V.\,Vuleti\'{c}, M.\,Lukin, J.\,Taylor, O.\,Firstenberg, T.\,Pohl, C.\,Murray, S.\,Hofferberth, D.\,Chang, N.\,Yao, and C.\,Nayak for discussions.
E.\,Z.~acknowledges funding from the European Research Council under the
European Union Seventh Framework Programme (FP/2007-2013) / European Research Counsel
Grant Agreement No.~306576 and the Carlsberg Foundation. M.\,J.\,G., M.\,F.\,M., and A.\,V.\,G.~thank the following funding
sources: Army Research Laboratory Center for Distributed Quantum Information, National Science Foundation Quantum Information Science program, Nation Science Foundation Physics Frontier Center at the Joint Quantum Institute, Air Force Office of Scientific Research, Army Research Office, and Army Research Office Multidisciplinary University Research Initiative.
M.\,F.\,M.~acknowledges start-up funding from Michigan State University.
W.\,H.\,P.\,Nielsen provided graphical assistance. 
\end{acknowledgments}

\bibliography{Rydberg}

\end{document}


\title{Supplemental Material for\\
``Correlated photon dynamics in dissipative Rydberg media''}

\author{Emil Zeuthen}
\email{zeuthen@nbi.ku.dk}
\affiliation{Niels Bohr Institute, University of Copenhagen, DK-2100 Copenhagen,
Denmark}
\affiliation{Institute for Theoretical Physics and Institute for Gravitational
Physics (Albert Einstein Institute), Leibniz Universit\"{a}t Hannover,
Callinstra{\ss}e 38, 30167 Hannover, Germany}

\author{Michael J. Gullans}
\affiliation{Joint Quantum Institute and Joint Center for Quantum Information
and Computer Science, National Institute of Standards and Technology
and University of Maryland, College Park, Maryland 20742, USA}
\author{Mohammad F. Maghrebi}
\affiliation{Joint Quantum Institute and Joint Center for Quantum Information
and Computer Science, National Institute of Standards and Technology
and University of Maryland, College Park, Maryland 20742, USA}
\affiliation{Department of Physics and Astronomy, Michigan State University, East Lansing, Michigan 48824, USA}
\author{Alexey V. Gorshkov}

\affiliation{Joint Quantum Institute and Joint Center for Quantum Information
and Computer Science, National Institute of Standards and Technology
and University of Maryland, College Park, Maryland 20742, USA}

\maketitle

\section{Successive projections of the polariton wave function\label{sec:Successive-projections_Appendix}}

In this Appendix, we comment further on the process by which the polariton wave train is generated in the medium. In particular, in Fig.~\ref{fig:Successive-projections}(b) of the main text, we assumed that the second scattering event occurs when the polariton has not passed the first blockade radius of the medium. In this Appendix, we comment on what happens if a scattering event occurs when the polariton straddles the rear of the first blockade radius of the medium.

\begin{figure}
\centering
\def\svgwidth{0.7\columnwidth}
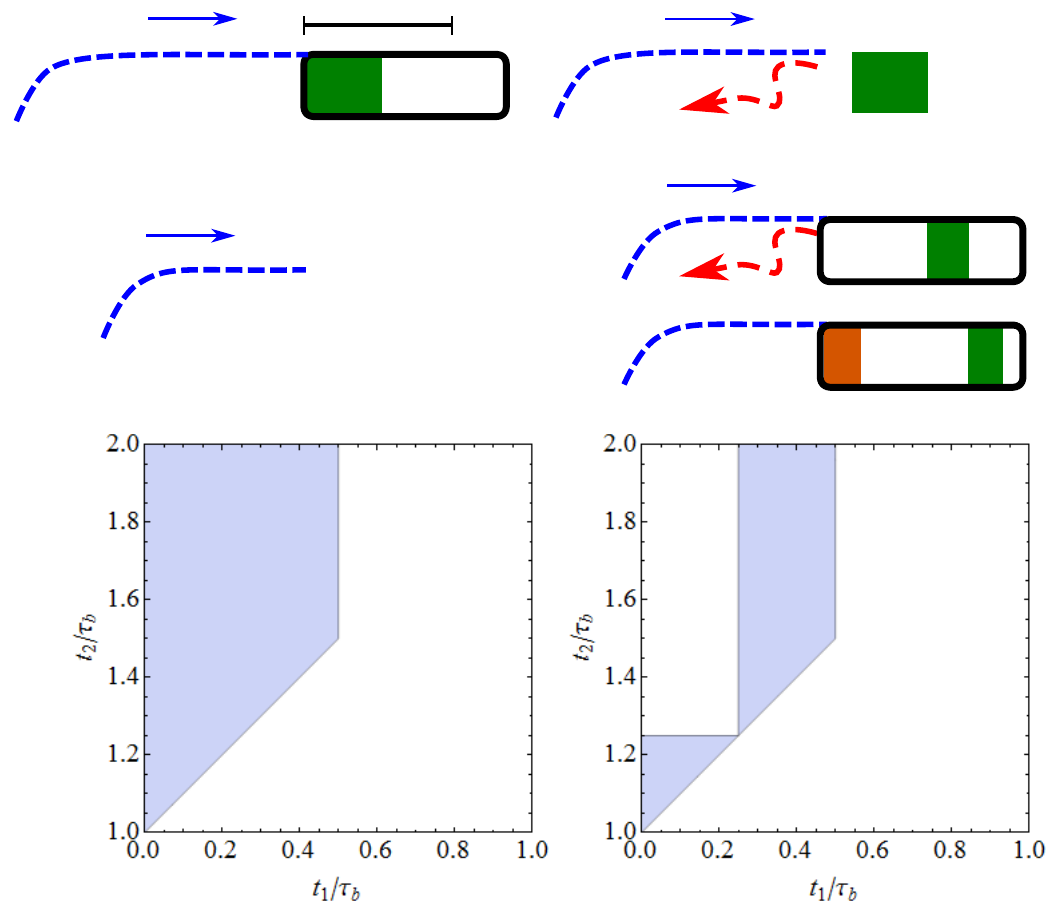

\caption{Formation of the polariton wave function by sequential projections
of an incoming square pulse according to the hard-sphere model in
the limit of perfect single-polariton EIT. (a) An incoming probe photon
is scattered (red dashed arrow) near the beginning of the medium thereby
projecting a polariton (green rectangle) in the medium. (b) The polariton
propagates further into the medium, and, at a subsequent time, a second
probe photon scatters, but since (in this instance) the polariton
could not have left the first $r_{\text{b}}$ of the medium, no additional
projection of the polariton wave function ensues. (c) The polariton
is now about to leave the first blockade radius of the medium, prompting
us to consider possible formation times $t_{2}$ of the second polariton
as described by the two-photon wave function in (c'), assuming the
pulse to arrive at $t=0$. The shaded region denotes the support of the wavefunction. (d) The first and second polaritons straddle
the rear and front boundaries of the first blockade radius as a scattering
event occurs; this causes a projection $\text{(c')}\rightarrow\text{(d')}$
on the two-body wave function producing a superposition state of the
first and second polariton [as indicated in (d')] being the scatterer.\label{fig:Successive-projections-Appendix}}
\end{figure}

As discussed in the main text, the temporal extent of the polariton
wave function is defined near the beginning of the medium [Fig.~\ref{fig:Successive-projections-Appendix}(a,b)].
Within the validity of the hard-sphere model, scattered photons are
ignorant as to the precise distance to the scatterer, since the scattering
event constitutes a projective, binary distance measurement. As a
consequence, when a polariton (whose temporal extent was defined near
the entrance) straddles the rear of the first blockade radius of the
medium [Fig.~\ref{fig:Successive-projections-Appendix}(c)], then a
scattering event cannot distinguish whether the scattering was caused
by this distant polariton, or whether the distant polariton had already
left and the scattering was instead caused by a newly formed polariton
near the entrance [see Fig.~\ref{fig:Successive-projections-Appendix}(d)].
The resulting projection caused by the scattering hence acts simultaneously
on the two polaritons leading to spatial entanglement [Fig.~\ref{fig:Successive-projections-Appendix}(d')]. 

As discussed in Appendix~\ref{sec:Determination-of-transmission_Appendix}, our model ignores such additional projections that can occur as the polariton leaves the first blockade radius of the medium. Since the rate of incoming photons determines both projection processes (the one occurring at the entrance into medium and the one occurring at the exit from the first blockade radius), we expect that ignoring the latter will not qualitatively affect our results.

\section{Determination of transmission function\label{sec:Determination-of-transmission_Appendix}}

In this Appendix, we present the details behind the derivation of the expressions for the average EIT transmission $\bar{\eta}_{\text{EIT}}(l)$
for a medium of length $l$ and the average effective blockade time
$\bar{\tau}_{\text{b}}$ given in Eqs.\ (\ref{eq:eta_EIT-bar}) and (\ref{eq:eta-tau-bar_result}) of the main text.


To estimate these quantities, we rely on the intuition presented in
Fig.~\ref{fig:Successive-projections}(a,b) of the main text: That
the temporal extent of the polariton wave function is determined near
the beginning of the medium by its first R-R scattering event, whereas
subsequent propagation within the first blockade radius is unaffected
by any additional R-R scattering events. This suggests a serialized
treatment of the R-R scattering and single-polariton EIT-filtering.
Applying this approach, we take the temporal extent $\tau$ of a polariton
to be defined by the first R-R event after its formation and assume
its EIT transmission probability $\eta_{\text{EIT}}(\tau,l)$ to be
a function only of $\tau$ and the propagation length $l$ in the
medium. In doing so, we ignore the additional projections that can
occur as the polariton leaves the first blockade radius of the medium,
as discussed in Appendix~\ref{sec:Successive-projections_Appendix}.

Averaging the EIT transmission probability $\eta_{\text{EIT}}(\tau,l)$
over the CW/Poisson distribution for the timing of the first R-R scattering
event $\tau$ amounts to
\begin{equation}
\bar{\eta}_{\text{EIT}}(l)=\langle\eta_{\text{EIT}}(\tau,l)\rangle_{\tau}=\int_{0}^{\infty}d\tau\mathcal{R}_{\text{in}}(\mathcal{R}_{\text{in}}\tau)e^{-\mathcal{R}_{\text{in}}\tau}\eta_{\text{EIT}}(\tau,l).\label{eq:eta-EIT-bar_average}
\end{equation}
From this average transmission probability, we estimate the average
effective blockade time as 
\begin{equation}
\bar{\tau}_{\text{b}}=\tau_{\text{b}}\int_{0}^{r_{\text{b}}}\frac{dl}{r_{\text{b}}}\bar{\eta}_{\text{EIT}}(l).\label{eq:tau-b_def_Appendix}
\end{equation}
Approximating $\eta_{\text{EIT}}(\tau,l)$ by the EIT transmission
of a square pulse subjected to Gaussian filtering,
\begin{equation}
\eta_{\text{EIT}}(\tau,l)=\int_{0}^{\tau}dt_{1}\int_{0}^{\tau}dt_{2}\frac{1}{\tau\sqrt{4\pi}\tau_{\text{EIT}}(l)}\exp\left[-\frac{(t_{1}-t_{2})^{2}}{4\tau_{\text{EIT}}^{2}(l)}\right]=\text{Erf}\left[\frac{\tau}{2\tau_{\text{EIT}}(l)}\right]-\frac{2\tau_{\text{EIT}}(l)}{\sqrt{\pi}\tau}\left(1-\exp\left[\frac{\tau^{2}}{4\tau_{\text{EIT}}^{2}(l)}\right]\right),
\end{equation}
the integrals in Eqs.~(\ref{eq:eta-EIT-bar_average},\ref{eq:tau-b_def_Appendix})
can be evaluated, yielding the expressions given in Eqs.\ (\ref{eq:eta_EIT-bar}) and (\ref{eq:eta-tau-bar_result}) of the main text.

\section{Hard-sphere correlation functions}

In this Appendix, we present the details behind the derivation of the first-order correlation function $G^{(1)}(\tau;\tau')$ presented in Eq.~(\ref{eq:G1-diag-Poisson}) of the main text, plot $G^{(1)}(\tau;\tau)$ for different parameters, and generalize the derivation to higher-order correlation functions. The derivations are carried out within the idealized hard-sphere dynamics for which single-polariton EIT effects are absent.

\subsection{First-order correlation function}

In this Section, we present the details behind the derivation of $G^{(1)}(\tau;\tau)$, Eq.~(\ref{eq:G1-diag-Poisson}) from the main text, and plot this function for different combinations of input rate and blockade time.

\begin{figure}
\begin{centering}
\includegraphics[width=0.7\columnwidth]{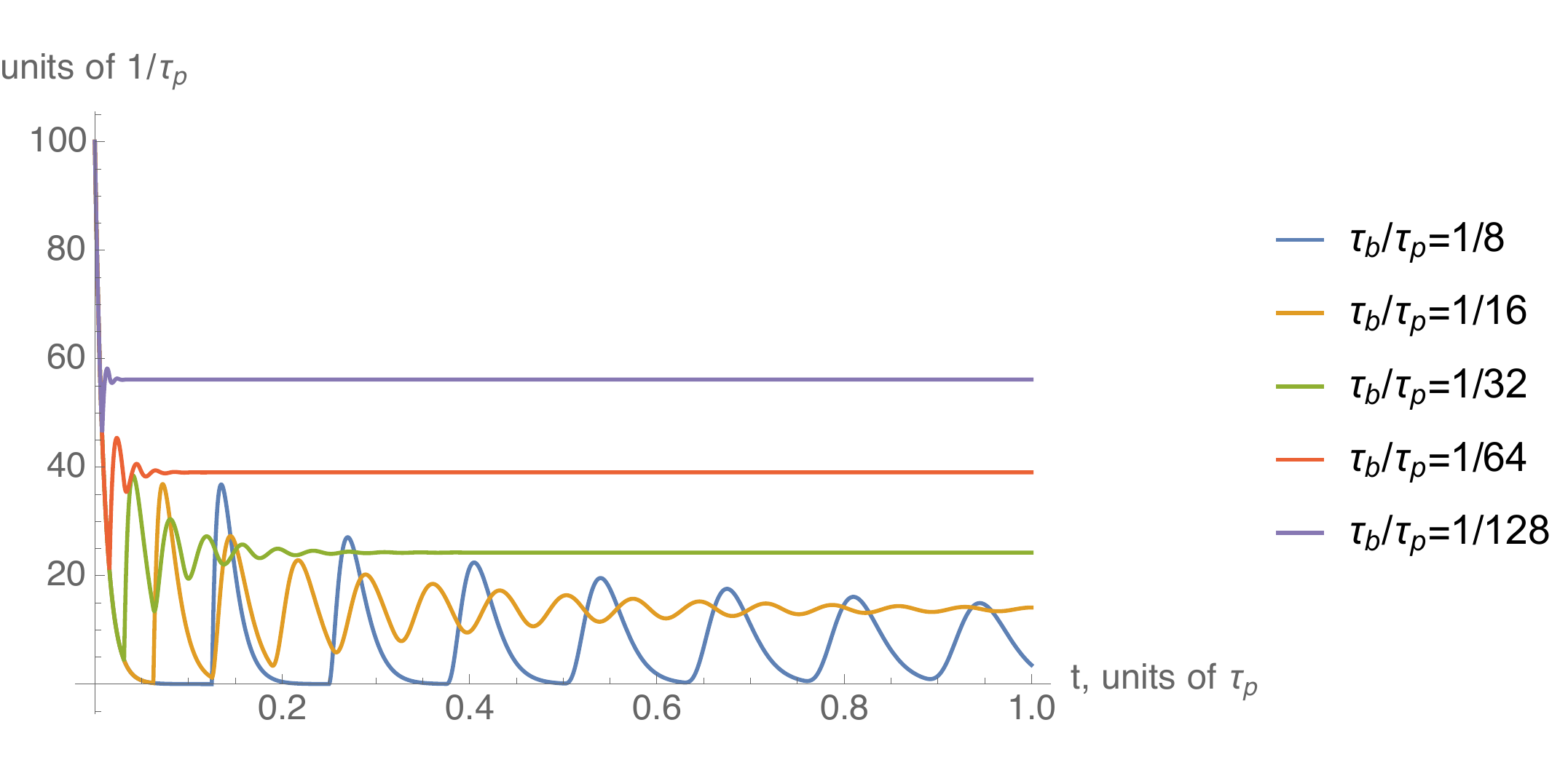}
\par\end{centering}
\caption{Ensemble-averaged output intensity $\langle\hat{I}(t)\rangle_{\hat{\rho}}\equiv G^{(1)}(t;t)$ according to the hard-sphere ansatz (ignoring EIT filtering).
Poisson-distributed input with fixed mean number of photons $\mathcal{R}_{\text{in}}\tau_{\text{p}}=100$
for different ratios of blockade time to pulse duration, $\tau_{\text{b}}/\tau_{\text{p}}$.\label{fig:G1-diag}}
\end{figure}

The same-time first-order correlation function $G^{(1)}(\tau;\tau)$ is given by the ensemble-averaged intensity profile $\langle\hat{I}(\tau)\rangle_{\hat{\rho}}=\langle\hat{\mathcal{E}}^{\dagger}(\tau)\hat{\mathcal{E}}(\tau)\rangle_{\hat{\rho}}=G^{(1)}(\tau;\tau)$,
where the subscript signifies expectation value with respect to the
density matrix $\hat{\rho}$ resulting from the hard-sphere interaction
(ignoring EIT filtering). For square-pulse Poisson input (of rate
$\mathcal{R}_{\text{in}}$), this function can be derived inductively
by propagating the initial condition that the medium is empty when
the input pulse arrives at the medium at time $\tau_{\text{s}}$ (and
using the fact that different segments of the input pulse are uncorrelated).
Let us first consider the probability density $P_{1}(t_{1}-\tau_{\text{s}})$
of the first Rydberg excitation occurring at a time $t_{1}\geq\tau_{\text{s}}$;
this is simply the product of the probability that no photons arrived
during the interval $[\tau_{\text{s}};t_{1}]$, i.e.~$\exp[-\mathcal{R}_{\text{in}}(t_{1}-\tau_{\text{s}})]$
for the Poisson distribution, and the arrival rate of photons $\mathcal{R}_{\text{in}}$,
so that we have
\begin{equation}
P_{1}(\tau)=\theta(\tau)\mathcal{R}_{\text{in}}\exp[-r\tau],\label{eq:P-1}
\end{equation}
where $\theta(\tau)$ is the Heaviside step function with the convention $\theta(0)=1$.
Next, let us construct the probability density $P_{2}(t_{2}-\tau_{\text{s}})$
that the second Rydberg excitation occurs at time $t_{2}$. Note that,
per the hard-sphere ansatz, this probability density can be non-zero
only for $t_{2}\geq\tau_{\text{s}}+\tau_{\text{b}}$. The \emph{conditional}
probability density of the second Rydberg excitation occurring at $t_{2}$ conditioned
on the first Rydberg excitation arriving at $t_{1}$ is just $P_{1}(t_{2}-t_{1}-\tau_{\text{b}})$,
where $P_{1}$ is given in Eq.~(\ref{eq:P-1}); i.e. the first Rydberg
excitation (at $t_{1}$) imposes an initial condition of an empty
medium at $t_{1}+\tau_{\text{b}}$ equivalent to the one at $\tau_{\text{s}}$.
The unconditional probability density $P_{2}(t_{2}-\tau_{\text{s}})$
for the arrival of the second Rydberg at $t_{2}$ is then found by
integrating the conditional density $P_{1}(t_{2}-t_{1}-\tau_{\text{b}})$
over $t_{1}\in[\tau_{\text{s}},t_{2}-\tau_{\text{b}}]$ weighted by
the probability density $P_{1}(t_{1}-\tau_{\text{s}})$ we found above
for $t_{1}$,
\begin{multline}
P_{2}(t_{2}-\tau_{\text{s}})=\theta(t_{2}-\tau_{\text{s}}-\tau_{\text{b}})\int_{\tau_{\text{s}}}^{t-\tau_{\text{b}}}dt_{1}P_{1}(t_{1}-\tau_{\text{s}})\times P_{1}(t_{2}-t_{1}-\tau_{\text{b}})\\
=\theta(t_{2}-\tau_{\text{s}}-\tau_{\text{b}})\mathcal{R}_{\text{in}}\exp[-\mathcal{R}_{\text{in}}(t_{2}-\tau_{\text{s}}-\tau_{\text{b}})]\times[\mathcal{R}_{\text{in}}(t_{2}-\tau_{\text{s}}-\tau_{\text{b}})].
\end{multline}
By iterating this argument, we find the probability density for the
arrival of the $R$'th Rydberg excitation at time $t_{R}$ to be given by (defining
$t_{0}\equiv\tau_{\text{s}}-\tau_{\text{b}}$ for convenience) 
\begin{equation}
P_{R}(t_{R}-\tau_{\text{s}})=\theta(t_{R}-\tau_{\text{s}}-(R-1)\tau_{\text{b}})\mathcal{R}_{\text{in}}e^{-\mathcal{R}_{\text{in}}(t_{R}-\tau_{\text{s}}-(R-1)\tau_{\text{b}})}\frac{[\mathcal{R}_{\text{in}}(t_{R}-\tau_{\text{s}}-(R-1)\tau_{\text{b}})]^{R-1}}{(R-1)!}.\label{eq:Arrival-prob-dens-R}
\end{equation}
$P_{R}(t_{R}-\tau_{\text{s}})$ in Eq.~(\ref{eq:Arrival-prob-dens-R}) 
is the probability density of the creation of a polariton at time
$t_{R}$ conditioned on $R-1$ polaritons having been created in the
preceding time interval $[\tau_{\text{s}},t_{R}-\tau_{\text{b}}]$.
This allows us to construct $G^{(1)}(\tau;\tau)$ simply by observing
that its value at a time $\tau=t-\tau_{\text{s}}$ after the onset
of the pulse only can have contributions from the first $\left\lceil (t-\tau_{\text{s}})/\tau_{\text{b}}\right\rceil $
polaritons created since $\tau_{\text{s}}$ per the hard-sphere ansatz;
summing these contributions, Eq.~(\ref{eq:Arrival-prob-dens-R}),
we find ($t\geq\tau_{\text{s}}$)
\begin{equation}
G^{(1)}(t-\tau_{\text{s}};t-\tau_{\text{s}})=\sum_{j=1}^{\left\lceil (t-\tau_{\text{s}})/\tau_{\text{b}}\right\rceil }P_{j}(t-\tau_{\text{s}})=\sum_{j'=0}^{\left\lfloor (t-\tau_{\text{s}})/\tau_{\text{b}}\right\rfloor }\mathcal{R}_{\text{in}}e^{-\mathcal{R}_{\text{in}}(t-\tau_{\text{s}}-j'\tau_{\text{b}})}\frac{[\mathcal{R}_{\text{in}}(t-\tau_{\text{s}}-j'\tau_{\text{b}})]^{j'}}{j'!},\label{eq:G1-diag-Poisson-Appendix}
\end{equation}
also given in Eq.\ (\ref{eq:G1-diag-Poisson}) of the main text (setting $\tau_{\text{s}}=0$ for simplicity).
We plot Eq.~(\ref{eq:G1-diag-Poisson-Appendix}) in Fig.~\ref{fig:G1-diag}
for different combinations of the input rate $\mathcal{R}_{\text{in}}$
and the blockade time in units of pulse duration $\tau_{\text{b}}/\tau_{\text{p}}$.
The width of the peaks are seen to increase with peak number while
their heights decrease. This is a symptom of the decay of the initial
condition of a vacant medium at $\tau_{\text{s}}$ when the pulse
arrives, corresponding to the decay of photon-photon correlations
in the output signal.

The off-diagonal elements $G^{(1)}(t-\tau_{\text{s}};t'-\tau_{\text{s}})$ can be expressed conveniently in terms of the diagonal elements (\ref{eq:G1-diag-Poisson-Appendix}) 
\begin{equation}
G^{(1)}(t-\tau_{\text{s}};t'-\tau_{\text{s}})=G^{(1)}(t_{<}-\tau_{\text{s}};t_{<}-\tau_{\text{s}})e^{-2\mathcal{R}_{\text{in}}\min\{|t-t'|,\tau_{\text{b}}\}}\times\begin{cases}
1, & \text{if } |t-t'|\leq\tau_{\text{b}}\\
\mathcal{R}_{\text{in}}^{-1}G^{(1)}(|t-t'|-\tau_{\text{b}};|t-t'|-\tau_{\text{b}}), & \text{if } |t-t'|>\tau_{\text{b}}
\end{cases},\label{eq:G1-general}
\end{equation}
where $t_{<} \equiv \min \{t,t'\}$.
Eq.~(\ref{eq:G1-general}) is derived in the main text for the special case $|t-t'|\leq\tau_{\mathrm{b}}$, see Eq.~(\ref{eq:G1_off-diag_coherent-ignore-edge}). In the case $|t-t'|>\tau_{\mathrm{b}}$ we must account for the fact that not all event histories for the intermediate time interval $[\min\{t,t'\}+\tau_{\text{b}};\max\{t,t'\}]$ are compatible with the medium being vacant at $\max\{t,t'\}$; hence we multiply by the probability that this is the case conditioned on the medium being vacant at $\min\{t,t'\}+\tau_{\text{b}}$, i.e., $\mathcal{R}_{\text{in}}^{-1}G^{(1)}(|t-t'|-\tau_{\text{b}};|t-t'|-\tau_{\text{b}})$.

\subsection{Higher-order correlation functions}

In this Section, we extend the arguments of the previous section to derive higher-order correlation functions, a possibility mentioned in conjunction with Eq.~(\ref{eq:G1-diag-Poisson}) of the main text.

The second-order correlation function, 
\begin{equation}
G^{(2)}(\tau_{1},\tau_{2};\tau_{2},\tau_{1})=\langle\dblcolon\hat{I}(\tau_{1})\hat{I}(\tau_{2})\dblcolon\rangle_{\hat{\rho}}=\langle\hat{\mathcal{E}}^{\dagger}(\tau_{1})\hat{\mathcal{E}}^{\dagger}(\tau_{2})\hat{\mathcal{E}}(\tau_{2})\hat{\mathcal{E}}(\tau_{1})\rangle_{\hat{\rho}},
\end{equation}
can be constructed from the diagonal elements $G^{(1)}(\tau;\tau)$
derived in Eq.~(\ref{eq:G1-diag-Poisson-Appendix}) by pursuing a similar logic:
As the product between the probability density of creating a polariton
at time $\tau_{\text{s}}+\min\{\tau_{1},\tau_{2}\}$ conditioned on
a vacant medium at $\tau_{\text{s}}$ and the probability density
of creating a polariton at time $\tau_{\text{s}}+\max\{\tau_{1},\tau_{2}\}$
conditioned on a polariton having been created at $\tau_{\text{s}}+\min\{\tau_{1},\tau_{2}\}$.
Importantly, the latter is independent of the event history of the
time interval $[\tau_{\text{s}};\tau_{\text{s}}+\min\{\tau_{1},\tau_{2}\}+\tau_{\text{b}}]$.
This is because the counting statistics of different time intervals
of the CW input are uncorrelated and conditioning on having a Rydberg
excitation created at $\tau_{\text{s}}+\min\{\tau_{1},\tau_{2}\}$
sets a boundary condition at $t'=\tau_{\text{s}}+\min\{\tau_{1},\tau_{2}\}+\tau_{\text{b}}$
equivalent to the initial condition at $\tau_{\text{s}}$ of a vacant
Rydberg medium. This argument leads to the expression:

\begin{equation}
G^{(2)}(\tau_{1},\tau_{2};\tau_{2},\tau_{1})=\Theta(|\tau_{2}-\tau_{1}|-\tau_{\text{b}})G^{(1)}(\min\{\tau_{1},\tau_{2}\};\min\{\tau_{1},\tau_{2}\})G^{(1)}(|\tau_{2}-\tau_{1}|-\tau_{\text{b}};|\tau_{2}-\tau_{1}|-\tau_{\text{b}}).\label{eq:G2_coherent-G1}
\end{equation}
This argument can be iterated to express the ``diagonal'' elements
($\tau_{i}=\tau_{i}'$) of the correlation function
\begin{equation}
G^{(N)}(\tau_{1},\ldots,\tau_{N};\tau_{N}',\ldots,\tau_{1}')\equiv\langle\dblcolon\prod_{i=1}^{N}\hat{\mathcal{E}}^{\dagger}(\tau_{i})\hat{\mathcal{E}}(\tau_{i}')\dblcolon\rangle_{\hat{\rho}}
\end{equation}
in terms of those of $G^{(1)}$ found in Eq.~(\ref{eq:G1-diag-Poisson-Appendix}).
Assuming a time-ordered set $\{\tau_{1},\ldots,\tau_{N}\}$, Eq.~(\ref{eq:G2_coherent-G1})
generalizes to (where $\tau_{0}\equiv-\tau_{\text{b}}$ for convenience)
\begin{equation}
G^{(N)}(\tau_{1},\ldots,\tau_{N};\tau_{N},\ldots,\tau_{1})=\prod_{i=1}^{N}\theta(\tau_{i}-\tau_{i-1}-\tau_{\text{b}})G^{(1)}(\tau_{i}-\tau_{i-1}-\tau_{\text{b}};\tau_{i}-\tau_{i-1}-\tau_{\text{b}}).\label{eq:G-N_from_G-1}
\end{equation}

\section{Comparison to numerical simulations of 3-level model for 2-photon input}

To check the serialization approximation for EIT filtering used when analyzing the scheme for generating single-photon trains, we
compare the results of this approximation against numerical simulations of the full set of equations of motion. The serialization approximation amounts to passing the state generated by the idealized R-R interaction [Fig.~\ref{fig:Successive-projections-Appendix}] through a linear EIT filter.

\begin{figure}
\begin{centering}
\begin{picture}(320,165) 
\put(0,0){\includegraphics[trim= 0 20 0 238, clip, width=0.5\columnwidth]{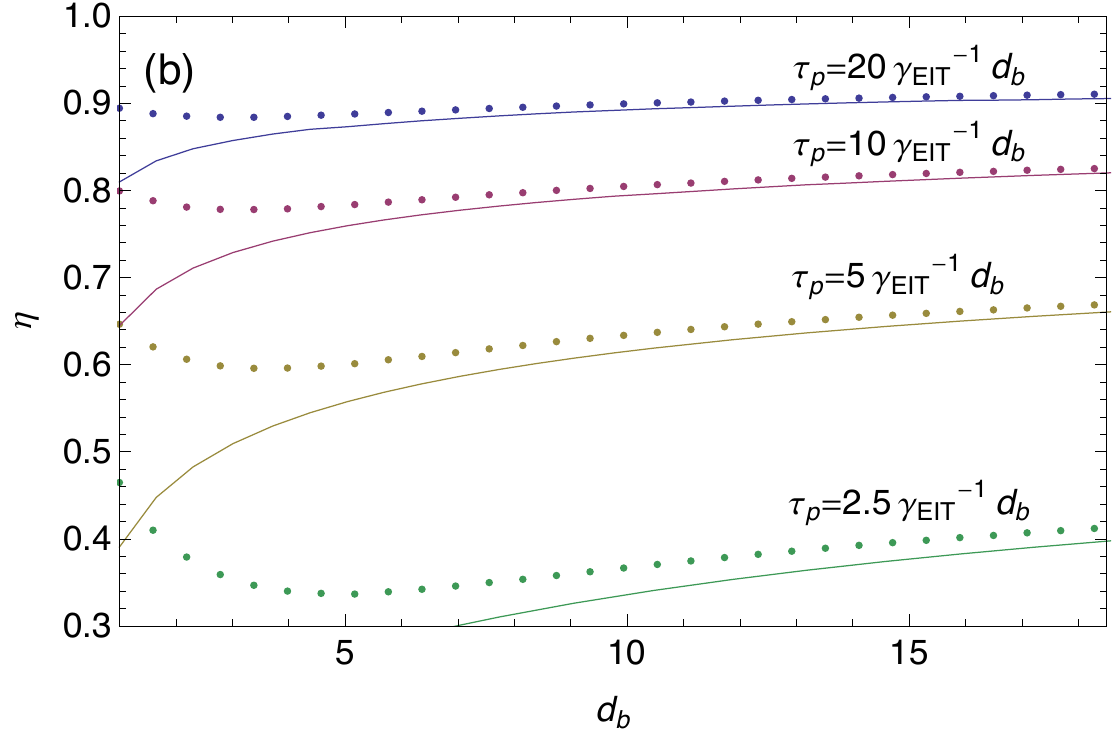}}
\color{white}{\put(40,154){\circle*{14}}}
\end{picture}
\par\end{centering}
\centering{}\caption{Transmission through a dissipative Rydberg-EIT medium. Comparison between full numerical simulation
and hard-sphere ansatz with post-EIT-filtering for the propagation
of a two-photon square pulse through a Rydberg medium of length $L=r_{\text{b}}$.\label{fig:R-out_serial-check}}
\end{figure}

For the comparison, we consider square-pulse two-photon Fock-state input with
 (which is reasonably feasible numerically). To
establish the prediction from our model, consider (for generality)
an arbitrary temporal pulse shape $h(t)$ which is non-zero only in
the time interval $[0;\tau_{\text{end}}\geq\tau_{\text{b}}]$ (normalized
as $\int h^{2}(t)dt=1$). According to the hard-sphere ansatz (ignoring
single-polariton EIT decay) the density matrix when the entire pulse
has entered the medium is
\begin{equation}
\hat{\rho}(t)=\overbrace{2\int_{0}^{\tau_{\text{end}}}d\tau_{1}\int_{\max\{\tau_{1}-\tau_{\text{b}},0\}}^{\tau_{1}}dt_{1}h^{2}(\tau_{1})h^{2}(t_{1})|\tilde{\psi}_{\tau_{1}}(t)\rangle\langle\tilde{\psi}_{\tau_{1}}(t)|}^{\text{one scattering event}}+\overbrace{2\int_{0}^{\tau_{\text{end}}-\tau_{\text{b}}}dt_{1}\int_{t_{1}+\tau_{\text{b}}}^{\tau_{\text{end}}}dt_{2}h^{2}(t_{1})h^{2}(t_{2})|\tilde{\psi}_{\emptyset}(t)\rangle\langle\tilde{\psi}_{\emptyset}(t)|}^{\text{no scattering event, }\emptyset},\label{eq:rho-2phot}
\end{equation}
where the normalized wave functions are
\begin{gather}
|\tilde{\psi}_{\tau_{1}}(t)\rangle=\frac{-\sqrt{v_{\text{g}}}}{\sqrt{\int_{\max\{\tau_{1}-\tau_{\text{b}},0\}}^{\tau_{1}}dt_{1}'h^{2}(t_{1}')}}\int_{\max\{\tau_{1}-\tau_{\text{b}},0\}}^{\tau_{1}}dt_{1}h(t_{1})\hat{S}^{\dagger}[v_{\text{g}}(t-t_{1})]|0\rangle, \label{eq:psi-tilde_tau1}\\
|\tilde{\psi}_{\emptyset}(t)\rangle=\frac{v_{\text{g}}}{\sqrt{\int_{0}^{\tau_{\text{end}}-\tau_{\text{b}}}dt_{1}'\int_{t_{1}'+\tau_{\text{b}}}^{\tau_{\text{end}}}dt_{2}'h^{2}(t_{1}')h^{2}(t_{2}')}}\int_{0}^{\tau_{\text{end}}-\tau_{\text{b}}}dt_{1}\int_{t_{1}+\tau_{\text{b}}}^{\tau_{\text{end}}}dt_{2}h(t_{1})h(t_{2})\hat{S}^{\dagger}[v_{\text{g}}(t-t_{1})]\hat{S}^{\dagger}[v_{\text{g}}(t-t_{2})]|0\rangle.\label{eq:psi-tilde_noscat}
\end{gather}
For our special case of a square pulse we have $h(t)=\tau_{\text{end}}^{-1/2}$, in which case the above integrals can be straightforwardly calculated.
We subject $\hat{\rho}$ 
given in Eq.\ (\ref{eq:rho-2phot}) to linear EIT filtering
corresponding to the full length $L$ of the Rydberg medium (taking
$L=r_{\text{b}}$). We focus on the transmission of the two-photon
component, i.e., the probability that neither R-R scattering nor EIT
decay occurs. 
This amounts to
filtering the second term in 
Eq.~(\ref{eq:rho-2phot}),
i.e., for square-pulse input, filtering the (unnormalized) wavefunction
of the state $(1-\tau_{\text{b}}/\tau_{\text{end}})|\tilde{\psi}_{\emptyset}(t)\rangle$, where $|\tilde{\psi}_{\emptyset}(t)\rangle$ is given in Eq.~(\ref{eq:psi-tilde_noscat}). The square norm of the filtered wavefunction
is the desired transmission probability. For simplicity, we approximated
the effect of a linear EIT medium by a Gaussian filter, yielding the
curves presented in Fig.~\ref{fig:R-out_serial-check}. This serialized approximation yields a pessimistic
estimate, since the sharp temporal features removed by the filter
are in general created somewhere in the interior of the medium thus
reducing the effective optical depth of the EIT-filtering effect.

The above theoretical prediction for square-pulse 2-photon Fock-state input is compared to  time-dependent numerical simulations of the pulse transmission through the medium using a three-level model for the atoms. The numerical methods are detailed in the supplementary material of Refs.~\cite{Peyronel2012,Maghrebi2015a}.
The comparison is shown in Fig.~\ref{fig:R-out_serial-check}, where we plot the transmission
of the two-photon component, showing good agreement for $d_{\text{b}}\apprge10$.

\section{Generation of trains of single photons from CW input}

In this Appendix, we present the details behind the derivation of the requirements that the input rate $\mathcal{R}_{\text{in}}$ has to satisfy in order to produce a regular train of single photons from CW input. Combining these requirements yield the scaling result presented as Eq.~(\ref{eq:N-max_result}) in the main text.

Since $G^{(N)}$ is simply related to $G^{(1)}$ by Eq.~(\ref{eq:G-N_from_G-1}), we will consider the signatures of regularity in $G^{(1)}$. Considering
the individual (unit-area) polariton peak profiles $P_{j}(t-\tau_{\text{s}})$
in Eq.~(\ref{eq:G1-diag-Poisson-Appendix}), the $j$'th peak is seen to be located
at $t_{j}=\tau_{\text{s}}+j(\tau_{\text{b}}+1/\mathcal{R}_{\text{in}})$
and hence the peak-to-peak separation is $\Delta t=\tau_{\text{b}}+1/\mathcal{R}_{\text{in}}$.
In the high-intensity limit, the peaks are well-separated, and we can
approximate Eq.~(\ref{eq:G1-diag-Poisson-Appendix}) by
\begin{equation}
\langle\hat{I}(\tau\sim t_{p}-\tau_{\text{s}})\rangle_{\hat{\rho}}\equiv G^{(1)}(\tau;\tau)\approx P_{p}(\tau),\label{eq:Int-crystal-approx}
\end{equation}
where $\tau\sim t_{p}-\tau_{\text{s}}$ means that $\tau$ is in the
neighborhood of the $p$'th peak at $t_{p}-\tau_{\text{p}}$ (or,
more precisely, that $p$ minimizes $|\tau-t_{p}+\tau_{\text{s}}|$). 

\subsection{Localization condition}

To derive a condition for well-separated peaks,  we consider the corrections
to Eq.~(\ref{eq:Int-crystal-approx}), which are simply the tails of
the other $P_{j}$ in Eq.~(\ref{eq:G1-diag-Poisson-Appendix}):
\begin{equation}
\langle\hat{I}(\tau\sim t_{p}-\tau_{\text{s}})\rangle_{\hat{\rho}}=P_{p}(\tau)+\sum_{j=1,j\neq p}^{\left\lceil \tau/\tau_{\text{b}}\right\rceil }P_{j}(\tau).
\end{equation}
As we shall see shortly, the width of $P_{p}(t)$ is sublinear in
$p$ ($\sim\sqrt{p}$) and hence grows slower than $t_{p}-\tau_{\text{s}}\propto p$.
For this reason, it is sufficient to ensure that each peak $P_{p}(t)$
is well-separated from its nearest neighbors. Thus, to have a train
of $N$ well-separated photons, we must ensure that the width $(\delta t)_{N}$
of the last peak $P_{N}$ is much less than the peak separation, $(\delta t)_{N}\ll\Delta t\approx\tau_{\text{b}}$
(in the high-intensity limit, $\mathcal{R_{\text{in}}}\gg1/\tau_{\text{b}}$).
Since the $p$'th peak width (HWHM) can be approximated for $p\gg1$
from Eq.~(\ref{eq:Arrival-prob-dens-R}) as 
\begin{equation}
(\delta t)_{p}\approx\frac{\sqrt{\ln(4)p}}{\mathcal{R}_{\text{in}}},\label{eq:Peak-width}
\end{equation}
this leads to the lower bound for $\mathcal{R}_{\text{in}}$ presented
in the main text (omitting the factor $\sqrt{\ln(4)}\sim1$ for simplicity).

\subsection{EIT transmittivity condition}
We now present the details behind the input rate requirement imposed by the finite EIT window. Using Eq.~(\ref{eq:G1_off-diag_coherent-ignore-edge}) of the main
text in the limit of well-separated peaks for which Eq.~(\ref{eq:Int-crystal-approx})
is applicable, we may filter the peaks individually
(here assuming $\tau_{\text{EIT}}\ll \tau_{\text{b}}$ and using the Gaussian EIT approximation),
\begin{equation}
\tilde{P}_{p}(\tau)\approx\int_{-\infty}^{\infty}dt_{1}\int_{-\infty}^{\infty}dt_{2}P_{p}(\min\{t_{1},t_{2}\})e^{-2\mathcal{R}_{\text{in}}|t_{1}-t_{2}|}\frac{1}{2\pi\tau_{\text{EIT}}^{2}}e^{-\frac{(t_{1}-\tau)^{2}}{2\tau_{\text{EIT}}^{2}}}e^{-\frac{(t_{2}-\tau)^{2}}{2\tau_{\text{EIT}}^{2}}}.\label{eq:P-tilde}
\end{equation}
Using Eq.~\eqref{eq:P-tilde} we estimate the single-polariton EIT transmission $\bar{\eta}_{\text{EIT}}$
as the integral over an individual filtered intensity peak $\tilde{P}_{p}(\tau)$ \footnote{The EIT loss determined here by post-filtering is found to be a factor
of 2 worse than $\bar{\eta}_{\text{EIT}}(r_{\text{b}})$ as given
by Eq.~(\ref{eq:eta-EIT-bar_average}) (comparing leading order terms in $\mathcal{R}_{\text{in}}\tau_{\text{EIT}}$). We ascribe this to the additional
projections that can occur when a polariton leaves the first $r_{\text{b}}$
of the medium [see Appendix \ref{sec:Successive-projections_Appendix} and  Fig.~\ref{fig:Successive-projections-Appendix}(d)].}, 
\begin{equation}
\bar{\eta}_{\text{EIT}}\equiv\int_{-\infty}^{\infty}d\tau\tilde{P}_{p}(\tau)=\exp\left([2\tau_{\text{EIT}}\mathcal{R}_{\text{in}}]^{2}\right)\text{erfc}\left(2\tau_{\text{EIT}}\mathcal{R}_{\text{in}}\right),\label{eq:eta-EIT_def}
\end{equation}
which is independent of the peak number $p$.
Expanding this in the limit $\tau_{\text{EIT}}\ll1/\mathcal{R}_{\text{in}}$
we find that
\begin{equation}
\bar{\eta}_{\text{EIT}}\approx1-\frac{4\mathcal{R}_{\text{in}}\tau_{\text{EIT}}}{\sqrt{\pi}}+\mathcal{O}[(\mathcal{R}_{\text{in}}\tau_{\text{EIT}})^{2}].
\end{equation}
Tolerating an EIT loss fraction of at most $\epsilon=1-\bar{\eta}_{\text{EIT}}$,
we are faced with an upper bound for $\mathcal{R}_{\text{in}}$,
\begin{equation}
\mathcal{R}_{\text{in}}\apprle\frac{\sqrt{\pi}}{4}\frac{\epsilon}{\tau_{\text{EIT}}},\label{eq:rate-req_EIT-Appendix}
\end{equation}
as presented in the main text (again omitting factors of order unity
for simplicity).

\bibliographystyle{apsrev4-1}
\bibliography{Rydberg}